\documentclass[10pt,conference]{IEEEtran}
\usepackage{amssymb,amsmath,amsfonts,bm}
\usepackage{graphicx}
\usepackage{subfigure}
\begin{document}

\title{Pseudo-codeword Landscape}

\author{
\authorblockN{Michael Chertkov}
\authorblockA{Theoretical Division and Center for Nonlinear
Studies\\ LANL, MS B213, T-13, Los Alamos, NM 87545\\ {\tt\small
chertkov@lanl.gov}} \and
\authorblockN{Mikhail Stepanov}
\authorblockA{Department of Mathematics,
The University of Arizona\\
617 N. Santa Rita Ave., Tucson, AZ 85721\\
{\tt\small stepanov@math.arizona.edu}}}

\maketitle

\begin{abstract}
We discuss the performance of Low-Density-Parity-Check (LDPC) codes
decoded by means of Linear Programming (LP) at moderate and large
Signal-to-Noise-Ratios (SNR). Utilizing a combination of the
previously introduced pseudo-codeword-search method and a new
``dendro" trick, which allows us to reduce the complexity of the LP
decoding, we analyze the dependence of the Frame-Error-Rate (FER) on
the SNR. Under Maximum-A-Posteriori (MAP) decoding the dendro-code,
having only checks with connectivity degree three, performs
identically to its original code with high-connectivity checks. For
a number of popular LDPC codes performing over the
Additive-White-Gaussian-Noise (AWGN) channel we found that either an
error-floor sets  at a relatively low SNR, or otherwise a transient
asymptote, characterized by a faster decay of FER with the SNR
increase, precedes the error-floor asymptote. We explain these
regimes in terms of the pseudo-codeword spectra of the codes.

\end{abstract}


\section{Introduction}

LDPC codes were introduced by Gallager \cite{63Gal} in anticipation
of the ease in their decoding.  The parity check matrix of an LDPC
code is sparse and the respective factor graph is locally tree like.
This suggests that the Belief Propagation (BP) decoding algorithm,
which would decode optimally in a loop free case, should also
perform well in the presence of relatively long loops. This
brilliant, but soon forgotten, guess of Gallager got a new life
after the discovery of the closely related turbo codes \cite{93BGT}
and later observations \cite{99Mac} that the LDPC codes may come
very close in their performance to the Shannon channel capacity
limit \cite{48Sha}. These considerations made LDPC codes top
candidates for emergent technologies in communications
\cite{802.11n} and data recording \cite{05KV}.

LP decoding of LDPC codes was introduced in \cite{05FWK} as a
relaxed, thus suboptimal but efficient, version of the optimal block
MAP decoding. The relation of LP decoding to the Bethe free energy
approach \cite{05YFW} and the BP equations and decoding was noticed
in \cite{05FWK},  and the point was elucidated further in
\cite{03KV,04VK,05VK,06CS,06CCc}. In short, LP may be considered as
the large SNR asymptotic limit of the BP solution,  where the latter
is interpreted as an extremum of the Bethe free energy functional.
(We will discuss this important relation below in Section
\ref{sec:LP}.) The big advantage of the LP comes from its discrete
nature and simplicity, leading in particular to the remarkable ML
certificate property \cite{05FWK}: if LP decodes to a codeword the
result is already optimal and cannot be improved as optimal
block-MAP would decode to the same codeword. Another useful
advantage offered by LP, in comparison with iterative BP (iterative
solution with a given schedule), is in a guaranteed convergence.
Finally, it is easy to implement the LP algorithm with general
purpose LP software. All these perks, however, do not come for free.
The main disadvantage of the LP is associated with a larger number
of degrees of freedom. The BP decoding operates in terms of
messages, totaling to twice the total number of edges in the Tanner
graph of the code (in the case of binary alphabet), while LP
decoding operates with local codewords, whose number grows
exponentially with the check degree, $q_\alpha$. Some number of
suggestions, briefly mentioned in the beginning of Section
\ref{sec:dendro}, were proposed to overcome the problem
\cite{05FWK,06TS,06VK}.

This paper suggests an alternative for reducing  the LP decoding
complexity. Our idea, explained in Section \ref{sec:dendro}, is to
change the graphical representation of the model by replacing all
checks of high degree by dendro-subgraphs (trees) with an
appropriate number of auxiliary checks of degree three and a number
of punctured, i.e. not transmitted, bits of degree two. We show that
the dendro-code and the original code have identical sets of
codewords and pseudo-codewords. Moreover,  for any configuration of
the channel output the results of MAP decodings are identical for
the two codes.

As shown in \cite{06CS} LP decoding allows a simple analysis of the
effective distance spectra of instantons, i.e. the most probable
erroneous configurations of the noise. The instantons are decoded
into pseudo-codewords, that typically are not codewords. The
pseudo-codeword-search method of \cite{06CS} suggests an efficient
algorithm for finding the pseudo-codewords with low effective
distance,  thus explaining the asymptotic behavior of FER in the
error-floor regime, i.e. at moderate and large SNRs.

Equipped with the new dendro-construction  we extend the
pseudo-codeword search algorithm and find the spectrum of the low
effective distance pseudo-codewords for codes which would otherwise
be impractical to decode by LP. The simulation results describing
the spectra are summarized in Section \ref{sec:spectra}. Here we
also report some results of Monte Carlo simulations. All together,
our simulations suggest that, error-floor performance wise, the
analyzed codes are split into roughly two qualitatively different
categories: either the error-floor sets in early, at relatively low
SNR, or otherwise, FER decay, with SNR increase, is steeper at
moderate SNR than at the SNR $\to\infty$ asymptote. We also give a
qualitative explanation of the phenomena.

\section{LP-decoding}
\label{sec:LP}

We consider a generic linear code, described by its parity check
$N\times M$ sparse matrix, $\hat{H}$, representing $N$ bits and $M$
checks. A codeword is a configuration, ${\bm\sigma}=\{\sigma_i=0,1|
i=1,\dots,N\}$, which satisfies all the check constraints: $\forall
\alpha=1,\dots,M$, $\sum_i H_{\alpha i}\sigma_i=0$ (mod~$2$). A
codeword sent through the channel is polluted and the task of
decoding becomes to restore the most probable pre-image of the
output sequence,  ${\bm x}=\{x_i\}$. The probability for
${\bm\sigma}$ to be a pre-image of ${\bm x}$ is
\begin{equation}
{\cal P}({\bm\sigma}|{\bm x})\!=\!Z^{-1}\prod_\alpha
\delta\biggl(\prod_{i\in\alpha}(-1)^{\sigma_i},1\biggr)
\exp\biggl(-\sum_ih_i\sigma_i\biggr), \label{Psx}
\end{equation}
where one writes $i\in\alpha$ if $H_{\alpha i}=1$; $Z$ is the
normalization coefficient (so-called partition function); the
Kronecker symbol, $\delta(x,y)$, is unity if $x=y$ and it is zero
otherwise; and ${\bm h}$ is the vector of log-likelihoods dependent
on the output vector ${\bm y}$. In the case of the AWGN channel with
the SNR ratio, $SNR=E_c/N_0=s^2$, bit transition probability is,
$\sim \exp(-2s^2(x_i-\sigma_i)^2)$, and the log-likelihood becomes,
$h_i=s^2(1-2x_i)$. The optimal block-MAP decoding maximizes ${\cal
P}({\bm\sigma}|{\bm x})$ over ${\bm\sigma}$. It can be restated as $
\arg\min_{{\bm\sigma}\in P}(\sum_ih_i\sigma_i)$, where $P$ is the
polytope spanned by the codewords \cite{05FWK}. Looking for
${\bm\sigma}$ in terms of a linear combination of all codewords of
the code, ${\bm \sigma}_v$: ${\bm\sigma}=\sum_v\lambda_v{\bm
\sigma}_v$, where $\lambda_v\geq 0$ and $\sum_v\lambda_v=1$, one
finds that block-MAP turns into a linear optimization problem. The
LP-decoding algorithm of \cite{05FWK} proposes to relax the
polytope, expressing ${\bm\sigma}$ in terms of a linear combination
of local codewords, i.e. codewords associated with single check
codes.

Prior to making a formal definition of LP decoding let us briefly
discuss its close relative, BP decoding \cite{63Gal,99Mac}. For an
idealized code on a tree, the BP algorithm is exactly equivalent to
the symbol-MAP decoding, which is reduced to the block-MAP (or
simply Maximum Likelihood, ML), in the asymptotic limit SNR
$\to\infty$. For any realistic code (with loops), the BP algorithm
is approximate, and it should actually be considered as an algorithm
solving iteratively certain nonlinear equations, called BP
equations. The BP equations are equations for extrema (e.g. minima
are of main interest) of the Bethe free energy \cite{05YFW}.
Minimizing the Bethe free energy, that is a nonlinear function of
the probabilities/beliefs, under the set of linear (compatibility
and normalizability) constraints, is generally a difficult task.

BP decoding turns into LP decoding at SNR $\to\infty$. In this
special limit, the entropy terms in the Bethe free energy can be
neglected and the problem turns to minimization of a linear
functional under a set of linear constraints. The similarity between
the LP and BP fixed points was first noticed in \cite{05FWK} and it
was also discussed in \cite{03KV,04VK,05VK,06CCc}. Stated in terms
of beliefs, LP decoding minimizes the self-energy,
\begin{eqnarray}
E=
\sum\limits_i\sum\limits_{\sigma_i} b_i(\sigma_i)h_i,\label{LP}
\end{eqnarray}
with respect to beliefs $b_i(\sigma_i)$,  which are defined as trial
probabilities for bit $i$ to be in the state $\sigma_i$. The beliefs
satisfy some equality and inequality constraints that allow
convenient reformulation in terms of a bigger set of beliefs defined
on checks, $b_\alpha({\bm\sigma}_\alpha)$, where, ${\bm
\sigma}_\alpha=\{\sigma_i|i\in\alpha,\sum_i H_{\alpha
i}\sigma_i=0\mbox{ (mod~$2$)}\}$, is a local codeword associated
with the check $\alpha$. The equality constraints are of two types,
normalization constraints (beliefs, as probabilities, should sum to
one) and compatibility constraints
\begin{equation}
\forall i,\ \forall\alpha\ni i:\
b_i(\sigma_i)=\sum\limits_{\sigma_\alpha\setminus\sigma_i}b_\alpha({\bm
\sigma}_\alpha),\
\sum\limits_{{\bm\sigma}_\alpha}b_\alpha({\bm\sigma}_\alpha)=1.\label{comp}
\end{equation}
Additionally, all the beliefs should be non-negative and smaller
than or equal to unity. This is the full definition of the LP
decoding. One can run it as is in terms of all the bit and check
beliefs, however it may also be useful to re-formulate the procedure
solely in terms of the bit beliefs. The ``small polytope"
formulation of LP is due to \cite{91Yan} and \cite{05FWK}.

\section{Dendro-LDPC}
\label{sec:dendro}

When it comes to decoding of the codes with high connectivity degree
of checks, $q_\alpha$, the most serious caveat of (otherwise simple
to state and analyze) LP decoding lies in its computational
complexity. Indeed, the number of check-related beliefs,
$b_\alpha({\bm\sigma}_\alpha)$, grows exponentially with $q_\alpha$,
$2^{q_\alpha-1}$,  thus making direct application of the powerful LP
machinery impractical for codes with large $q_\alpha$.

\begin{figure} [b]
\includegraphics[width=0.4\textwidth]{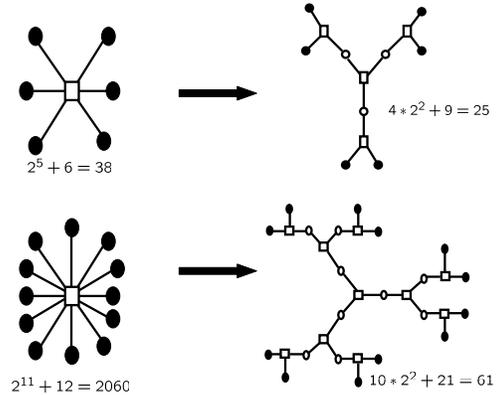}
\caption{Example of the dendro operation on single-check codes.
Numbers count beliefs (degrees of freedom) required for LP
evaluation.} \label{fig:dendro}
\end{figure}

However, many of the constraints associated with the check beliefs
are not really required for decoding. It was argued in \cite{05FWK}
that the number of useful constraints per check can be reduced to
$\sim q_\alpha$. This result was improved in \cite{06TS}, where an
impressive $O(1)$, thus $q_\alpha$-independent, scaling in the
number of required constraints was experimentally achieved by
adaptive scheduling of constraints and early termination in the case
of successful decoding. (Note  that the number of log-likelihood
dependent scheduling operations required here is $\sim q_\alpha$,
thus complexity of the entire algorithm is linear in $q_\alpha$.)
Armed with the observations of \cite{05FWK,06TS} and also of
\cite{06VK}, where a BP-style relaxation of LP achieving overall
linear scaling in $q_\alpha$ was proposed,  we extend the list of
useful tricks by the dendro scheme explained below. The scheme,
demonstrating overall linear scaling in $q_{\alpha}$, does not
require log-likelihood dependent adaptation.

Our strategy in dealing with the checks of high degree is through
modification of the graphical model (Tanner graph) of the code. We
simply replace the check by a dendro, i.e. tree,  graph with the
same number of leaves as the number of bit neighbors, $q_\alpha$, in
the original graph. All bits inside the dendro construction,  i.e.
these that are not leaves,  are the auxiliary, punctured, bits.  The
new dendro-checks are all of degree three, while the punctured bits
are all of degree two. The punctuated bits are not transmitted, thus
the log-likelihoods at the bits are zeros. The construction is
illustrated in Fig.~\ref{fig:dendro}.

The simple dendro construction is advantageous for decoding as the
total number of beliefs is seriously reduced. It becomes linear in
$q_\alpha$ at $q_\alpha\to\infty$ for the dendro-code correspondent
to a single-check code, as opposed to $2^{q_\alpha}$ for the
original code.

It is straightforward to verify that the codewords of the original
code and of the dendro-code are in one-to-one correspondence.
Indeed, the codewords of a code are controlled by checks, formally
expressed in terms of the product of the Kronecker symbols in
Eq.~(\ref{Psx}). On the other hand, any check constraint can be
explicitly rewritten as
\begin{eqnarray}
 && \delta\biggl(\prod_{i\in\alpha}(-1)^{\sigma_i},1\biggr)=
 \sum_{{\bm\sigma}^{(pun)}_\alpha}\prod_{\beta\in d(\alpha)}\delta\biggl(
 \prod_{j\in\beta}(-1)^{\sigma_j},1\biggr)\nonumber\\ &&
 \times\prod_{\beta\in\partial(\alpha)}\delta\biggl(
 \prod_{j\in\beta}(-1)^{\sigma_j},\prod_{i\in\beta}(-1)^{\sigma_i}\biggr),
 \label{dendro}
\end{eqnarray}
where $d(\alpha)$[$\partial(\alpha)$] are the sets of dendro checks
replacing check $\alpha$ such that the checks neighbor only [not
only] punctured bits;  and ${\bm\sigma}^{(pun)}_\alpha$ is the
vector of punctured bits originating from the check, $\alpha$, of
the original code. The lhs and rhs of Eq.~(\ref{dendro}) correspond
to the check constraints of the original code and the dendro code
respectively.  Putting it in a less formal way,  once the values of
the bits of the original codes are known, the punctured bits of the
respective dendro code are unambiguously restored. Furthermore,
since the punctuated bits are not transmitted and have zero
log-likelihoods, one finds that MAP decoding of the original code
and of its dendro counterpart generate exactly the same results.

Comparing the LP decoding of the two codes,  it is useful to turn to
the notion of the graph covers discussed in \cite{03KV,04VK,05VK}.
The pseudo-codewords of an LDPC code are in one-to-one
correspondence with the codewords of the respective family of
graph-cover LDPC codes. The graph covers are constructed by
replicating the total number of checks and nodes of the code by the
same positive integer,  the cover degree, and by connecting the
bits/checks with replicas of their original neighbors. The family of
graph covers can be generated both for the original code and for the
respective dendro code. The number of graph covers of the same
degree is larger for the dendro-code then for its original code.
More specifically,  for each standard-cover one gets a family of
equivalent dendro-covers. Each dendro-cover from the family will get
exactly the same set of codewords as of the original code.  This
statement follows directly from the  previous paragraph. Therefore,
the set of pseudo-codewords,  understood as the codewords of the set
of covers, will be exactly the same for the original code and its
dendro-counterpart. Let us notice that this statement does not
necessarily mean that decoding of the same output configuration by
the two codes will necessarily give the same result.

\section{Error-floor and Pseudo-codeword-search algorithm \cite{06CS}}
\label{sec:PCS}

If the LP decoding does not decode to a correct codeword then it
usually yields a non-codeword pseudo-codeword, which is a special
configuration of beliefs containing some rational values
\cite{05FWK,04VK}. An important characteristic of the decoding
performance is the Frame Error Rate (FER) calculating the
probability of a decoding failure. FER decreases as SNR increases.
The form of this dependence gives an ultimate description of the
coding performance. Any decoding to a non-codeword pseudo-codeword
is a failure. Decoding to a codeword can also be a failure, which
counts as a failure under the ML decoding. For large SNR, splitting
of the two (FER vs SNR) curves, representing ML decoding and
approximate decoding (say LP decoding) is due to the
pseudo-codewords. The actual asymptotics of the two curves for the
AWGN channel at the largest SNRs,  in the so-called error-floor
domain, are $\mbox{FER}_{\mbox{\scriptsize ML}}\sim \exp(-
d_{\mbox{\scriptsize ML}}\cdot s^2/2)$ and
$\mbox{FER}_{\mbox{\scriptsize LP}}\sim \exp(-d_{\mbox{\scriptsize
LP}}\cdot s^2/2)$, where $d_{\mbox{\scriptsize ML}}$ is the Hamming
distance of the code and the $d_{\mbox{\scriptsize LP}}$ is the
effective distance of the code, specific for the LP decoding. The LP
asymptotic is normally shallower than the one of MAP,
$d_{\mbox{\scriptsize LP}}<d_{\mbox{\scriptsize ML}}$. The error
floor can start or change its behavior at values of FER inaccessible
by Monte-Carlo simulations. This fact emphasizes the importance of
the pseudo-codewords analysis \cite{04Ric}.

For a generic binary linear code that is used for data communication
over binary-input output-symmetric channel, it is easy to show that
FER is invariant under change of the original codeword (sent into
the channel). Therefore, for the purpose of FER evaluation, it is
sufficient to analyze  statistics exclusively for the case of one
known original codeword,  and the choice of zero codeword is
natural. Then, calculating the effective distance of a code, one
makes an assumption that there exists a special configuration (or a
few special configurations) of the noise, instantons according to
the terminology of \cite{05SCCV}, describing the large SNR
error-floor asymptotic for FER. Suppose a pseudo codeword,
$\tilde{\bm \sigma}=\{\tilde{\sigma_i}=b_i(1);\ i=1,\dots,N\}$,
corresponding to the most damaging configuration of the noise
(instanton), ${\bm x}_{\mbox{\scriptsize inst}}$, is found. Then
finding the instanton configuration itself (i.e. respective
configuration of the noise) is not a problem, one only needs to
maximize the transition probability with respect to the noise field,
${\bm x}$, taken at ${\bm \sigma}=0$ under the condition that the
self-energy calculated for the pseudo-codeword in the given noise
field ${\bm x}$ is zero (i.e. equal to the value of the self energy
for the zero code word). The resulting expression for the optimal
configuration of the noise (instanton) in the case of the AWGN
channel is
 ${\bm x}_{\mbox{\scriptsize inst}}=(\tilde{\bm
 \sigma}\sum_i\tilde{\sigma}_i)/(2\sum_i\tilde{\sigma}_i^2)$,
and the respective effective distance is
 $d_{\mbox{\scriptsize LP}}=(\sum_i\tilde{\sigma}_i)^2/
 \sum_i\tilde{\sigma}_i^2$.
This definition of the effective distance was first described in
\cite{01FKKR},  with the first applications of this formula to LP
decoding discussed in \cite{03KV} and \cite{05VK}. Note also that
the expressions are reminiscent of the formulas derived by Wiberg
and co-authors in \cite{95WLK} and \cite{96Wib}, in the context of
the computational tree analysis applied to iterative decoding with a
finite number of iterations.

Let us now describe the pseudo-codeword-search algorithm, introduced
in \cite{06CS}. {\bf Start:} Initiate a starting configuration of
the noise, $ {\bm x}^{(0)}$. Noise is counted from the zero codeword
and it should be sufficiently large to guarantee convergence of LP
to a pseudo-codeword different from the zero codeword. {\bf Step 1:}
The LP decodes ${\bm \sigma}^{(k)}$ to $\{b_i^{(\mbox{\scriptsize
LP},k)}(\sigma_i),b_\alpha^{(\mbox{\scriptsize
LP},k)}(\sigma_\alpha)\}$. {\bf Step 2:} Find ${\bm y}^{(k)}$, the
weighted median in the noise space between the pseudo codeword,
${\bm \sigma}^{(k)}$, and the zero codeword.  The AWGN expression
for the weighted median is
 ${\bm y}^{(k)}=({\bm \sigma}^{(k)}\sum_i\sigma_i^{(k)})/(2
 \sum_i\big(\sigma_i^{(k)}\big)^2)$. {\bf Step 3:}
 If ${\bm y}^{(k)}={\bm y}^{(k-1)}$, then $k_*=k$ and the algorithm
terminates. Otherwise go to Step 2, assigning ${\bm x}^{(k+1)}={\bm
y}^{(k)}+\varepsilon$ for some very small $\varepsilon$.
($+\varepsilon$ prevents decoding into the zero codeword, keeping
the result of decoding within the erroneous domain.) {\bf Output}
configuration ${\bm y}^{(k_*)}$ is the configuration of the noise
that belongs to the error-surface surrounding the zero codeword.
(The error-surface separates the domain of correct LP decisions from
the domain of incorrect LP decisions.) Moreover, locally,  i.e. for
the given part of the error-surface equidistant from the zero
codeword and the pseudo codeword ${\bm \sigma}^{(k_*)}$, ${\bm
y}^{(k_*)}$ is the nearest point of the error-surface to the zero
codeword.

We repeat the algorithm many times, picking the initial noise
configuration randomly,  however guaranteeing that it would be
sufficiently far from the zero codeword so that the result of the LP
decoding (first step of the algorithm) is a pseudo-codeword distinct
from the zero codeword. We showed in \cite{06CS} that the algorithm
converges,  and that it does so in a relatively small number of
iterations.  The error-floor performance of the coding scheme is
characterized by the spectra of the effective distances derived over
multiple evaluations of the pseudo-codeword-search algorithm.

We can easily extend the pseudo-codeword-search algorithm to the
dendro-LDPC codes decoded by LP. The dendro version of the algorithm
is actually identical to the one described above with the exception
of how the punctured nodes are treated. First, one should always
zero the log-likelihoods at all the punctured nodes and,  second,
calculating the weighted medians, one should exclude punctured nodes
from the sum.

\section{Pseudo-codeword spectra: Results and Analysis}
\label{sec:spectra}

\begin{figure}[t]

\subfigure{
\includegraphics[width=0.23\textwidth]{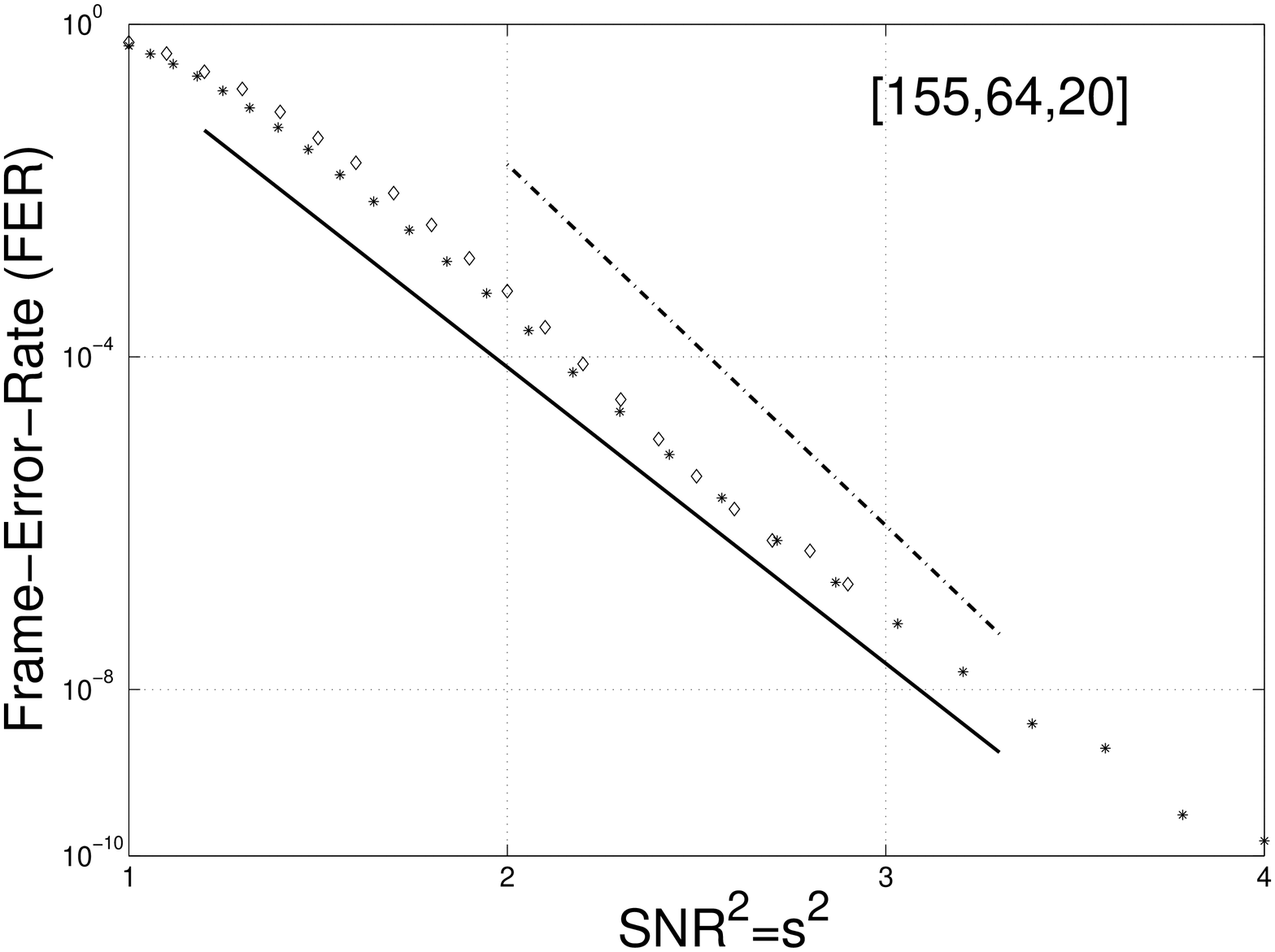}
\includegraphics[width=0.23\textwidth]{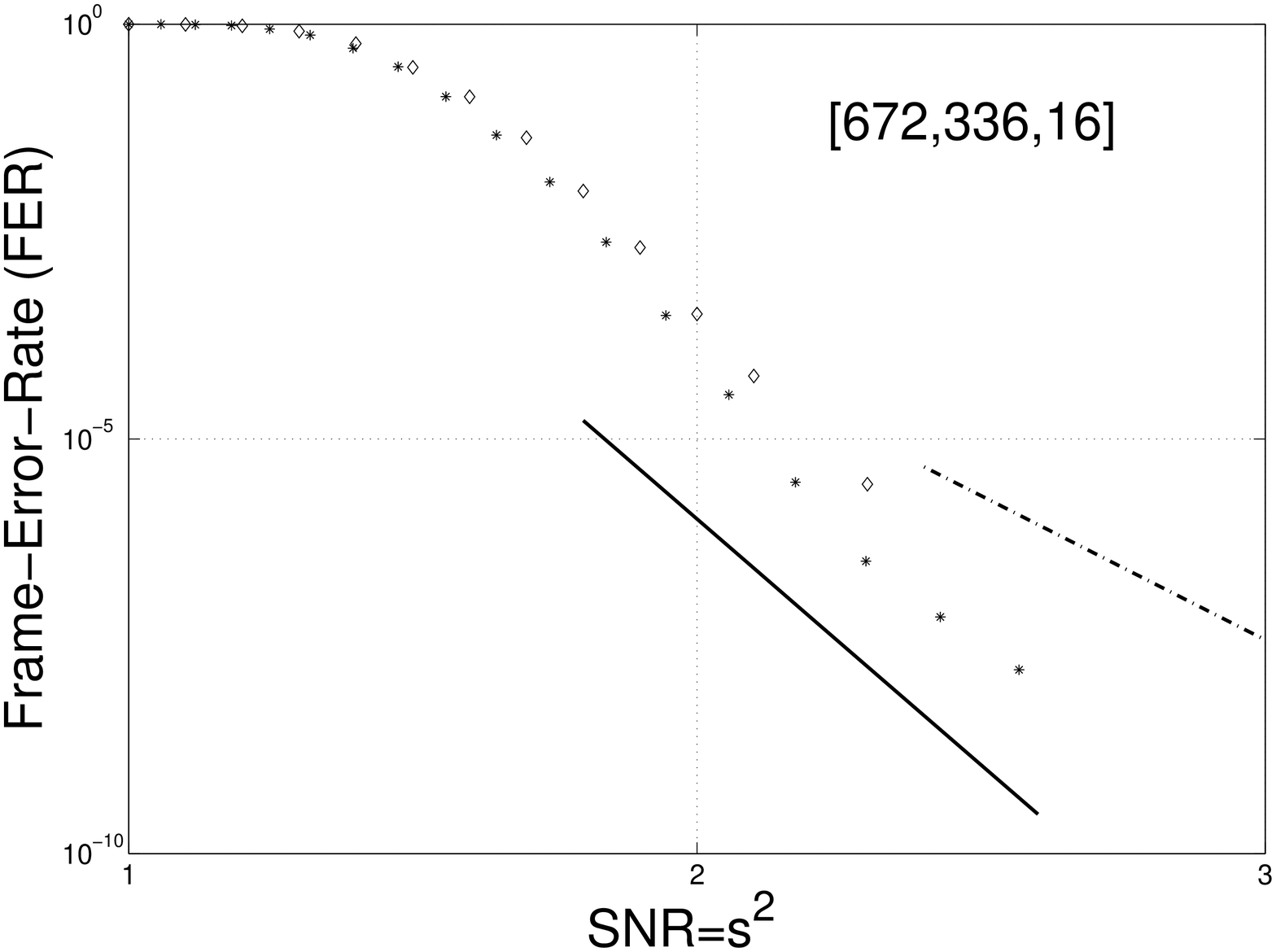}
 }

\subfigure{
\includegraphics[width=0.23\textwidth]{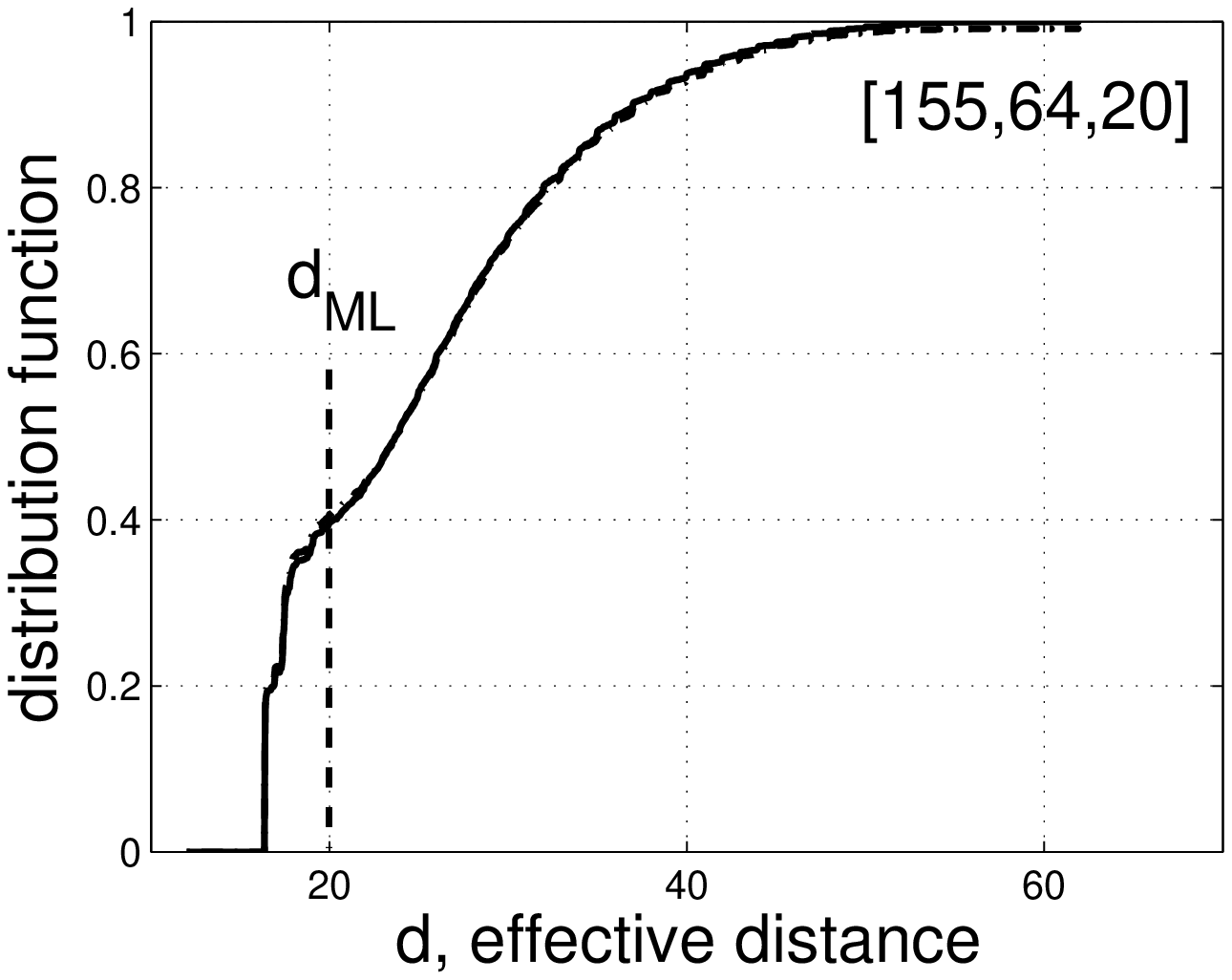}
\includegraphics[width=0.23\textwidth]{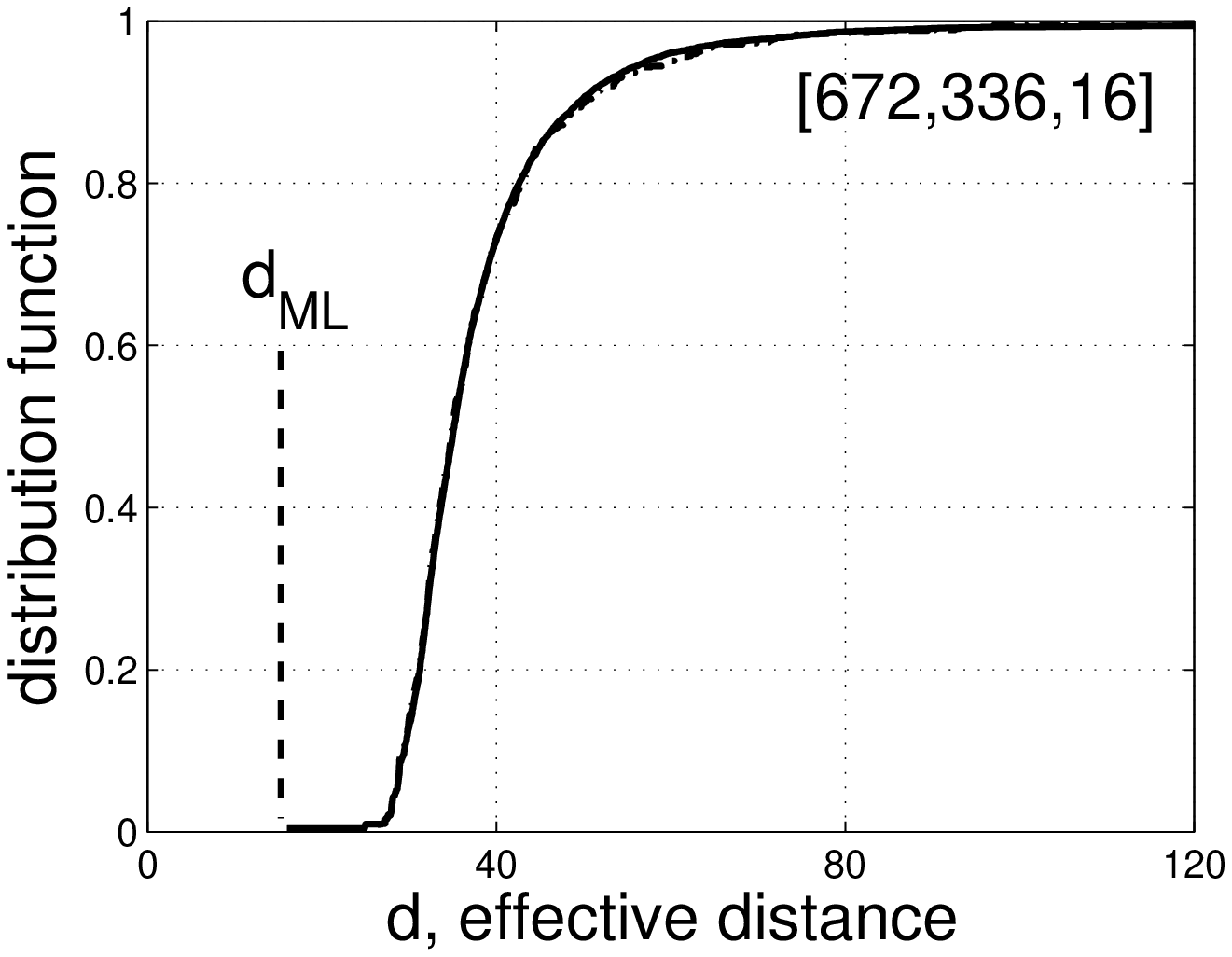}
 }

\subfigure{
\includegraphics[width=0.23\textwidth]{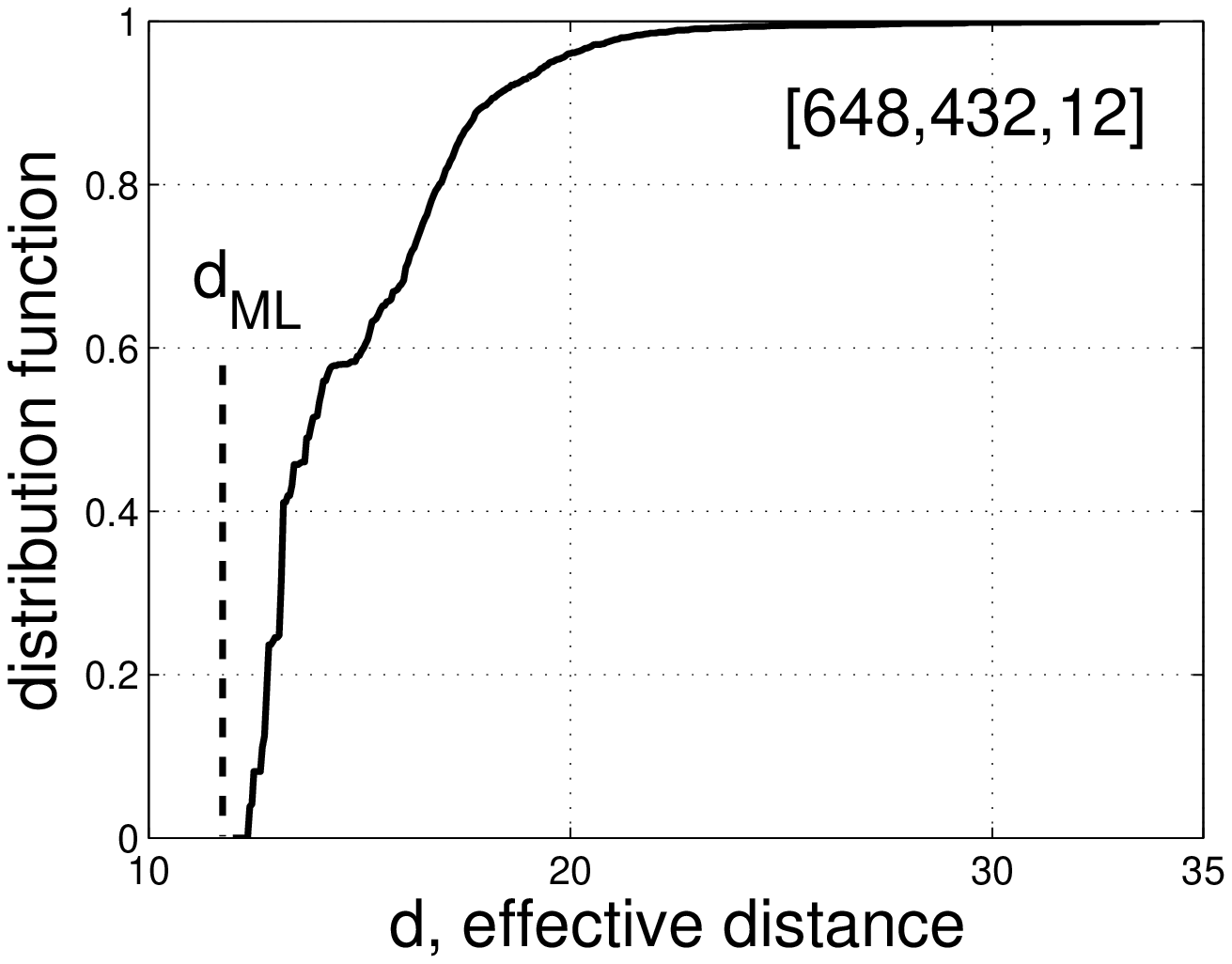}
\includegraphics[width=0.23\textwidth]{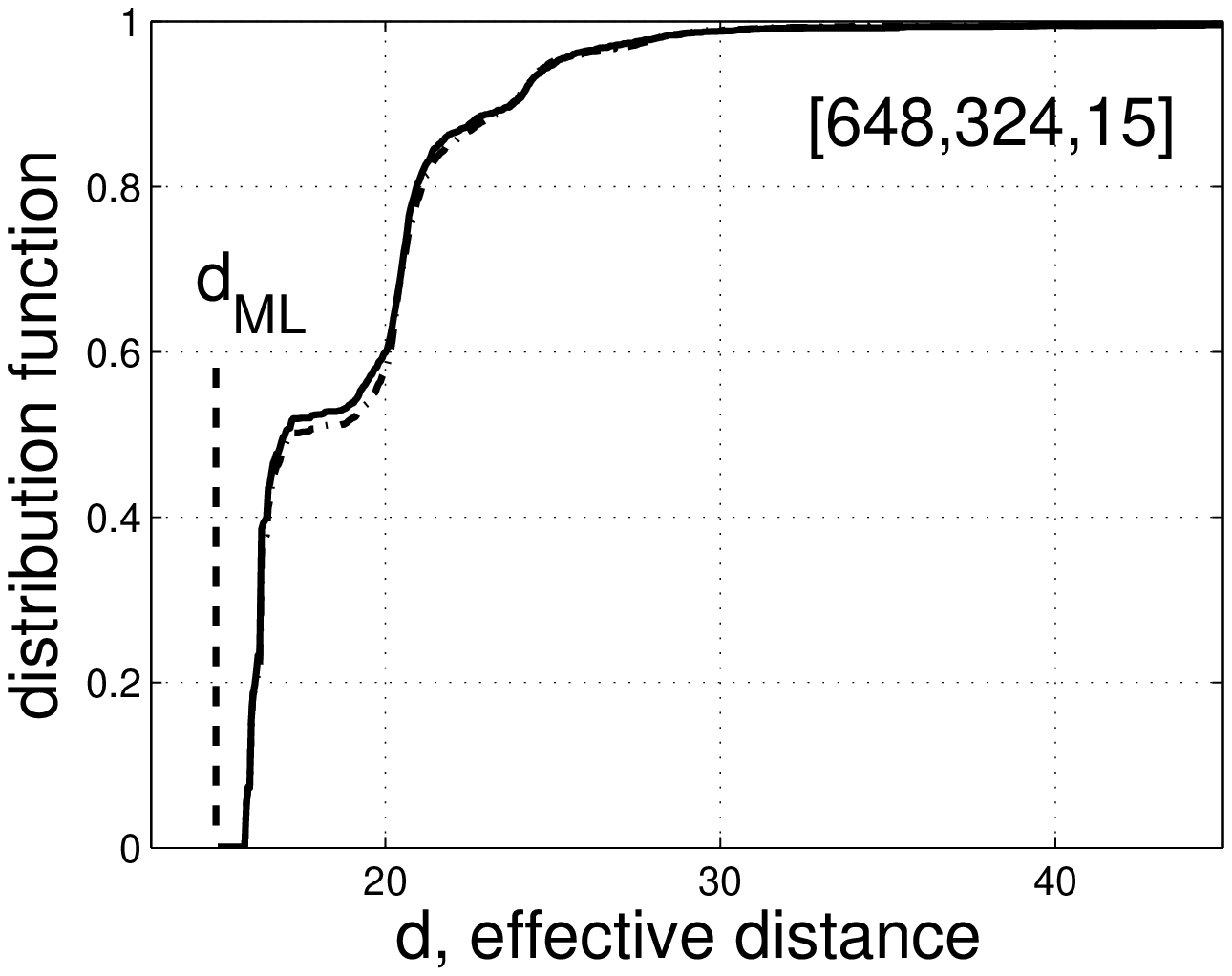}}

\subfigure{
\includegraphics[width=0.23\textwidth]{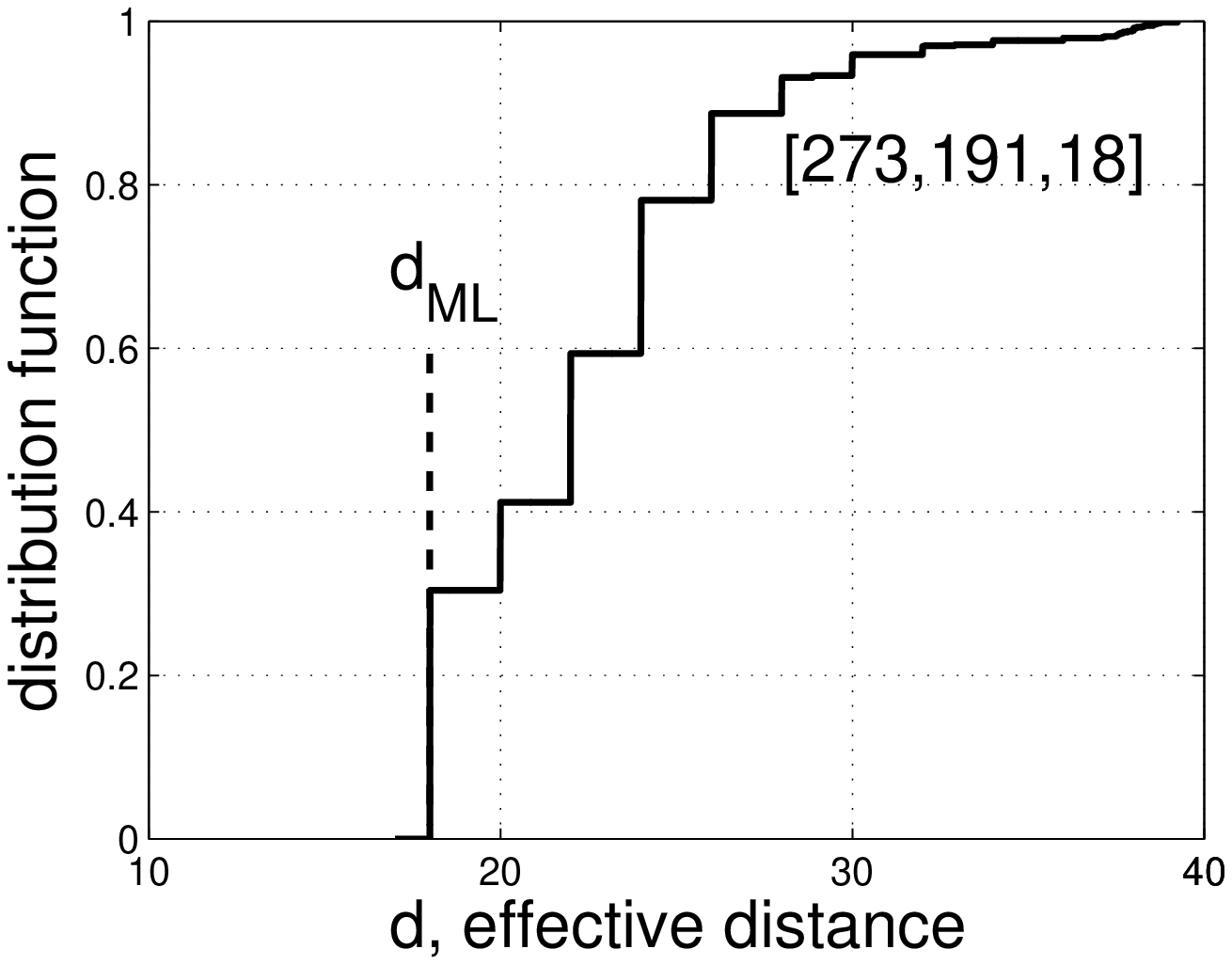}
\includegraphics[width=0.23\textwidth]{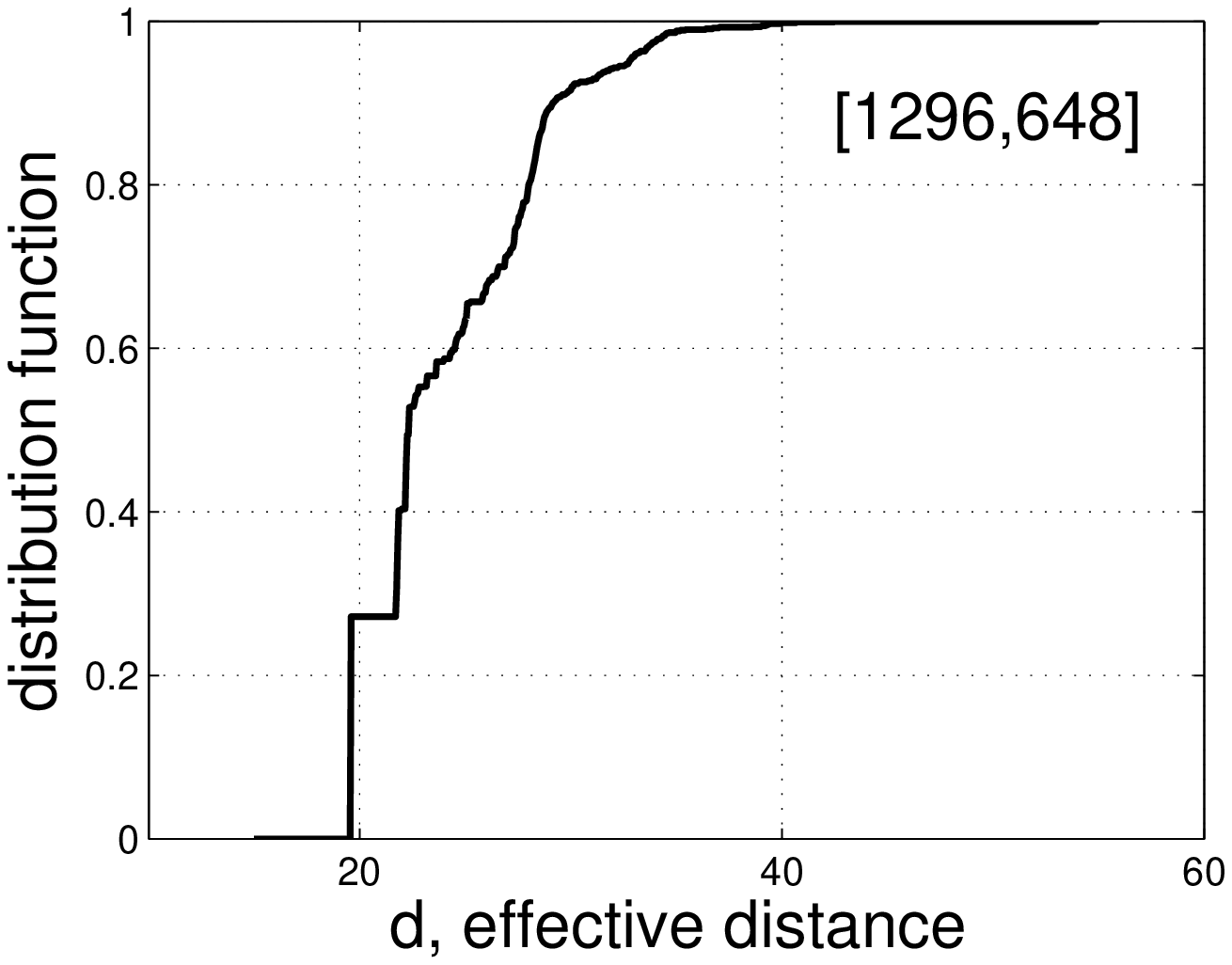}
}

\caption{The two plots from the first row show results of
Monte-Carlo simulations for the $[155,64,20]$ and $[672,336,16]$
codes. Stars and diamonds stand for BP (1024 iterations) and LP
decodings respectively. Straight and dashed lines mark the
asymptotic controlled by the pseudo-codewords with the lowest
effective weight and the MAP asymptotic respectively. The six
remaining plots from the three lower rows show probability density
function of the effective distance (that we also refer to as
frequency spectra) for the six codes analyzed. Solid and dashed
curves (almost coinciding, thus difficult to distinguish) correspond
to the dendro-codes and the original codes respectively. We indicate
position of the respective Hamming distance by a marker,  whenever
it is known.} \label{fig:MC+dens}
\end{figure}

We experimented with the $[155,64,20]$ code \cite{01TSF}, Margulis
$p=7$ code $[672,336,16]$ \cite{82Mar}; $[648,324,15]$,
$[648,432,12]$ and $[1296,648]$ codes from the 802.11n draft
\cite{802.11n}, and the $[273,191,18]$ projective geometry code. The
results are shown in Fig.~(\ref{fig:MC+dens}). Dendro counterparts
were generated for all the codes. For the dendro-codes, and whenever
feasible for the original codes, we have found the frequency spectra
of the pseudo-codewords by the method explained in Section
\ref{sec:PCS}. We experimentally confirmed the prediction of Section
\ref{sec:dendro}, that the set of pseudo-codewords of the original
codes and respective dendro-codes are identical. Moreover, we found
that the corresponding pseudo-codeword spectra are almost
indistinguishable (up to variation in the number of samples) from
each other. For the first two codes from the list, we also performed
direct Monte-Carlo simulations.

The rest of the manuscript contains discussion of the results.
Comparing  $[155,64,20]$ and $[672,336,16]$ codes we conclude that
the two codes demonstrate qualitatively different features, that are
consistently seen both in the MC simulations and the
pseudo-codewords frequency spectra.

In the case of the $[155,64,20]$ code, the pseudo-codeword spectrum
starts form $d_{min}\approx 16.404$ and grows continuously to the
higher values,  e.g. passing though $d_{ML}=20$ without any visible
anomaly. The growth, starting immediately from $d_{min}$, is fast,
indicating that the frequency of the low-effective distance
configurations is considerable, i.e. $O(1)$. This form of the
pseudo-codeword spectra is consistent with what is seen in the MC
simulations: the error-floor asymptotic of FER, $\sim
\exp(-d_{min}s^2/2)$, corresponding to the pseudo-codeword with the
lowest effective weight, sets in early.

The behavior demonstrated by the $[672,336,16]$ code is different.
Looking, first, at the pseudo-codeword spectra, we find that the
configuration with the lowest effective distance is actually a
codeword with $d_{ML}=16$. We also find in the spectrum two other
codewords corresponding to $d=24$ and $d=25$. Even though the low
distance codewords were observed, their frequencies were orders of
magnitude smaller then of other pseudo-codeword configurations found
at $d\gtrsim 27.33$. Emergence of the gap suggests that in spite of
the fact that the relatively small Hamming distance describes the
largest SNR asymptotic of FER, the moderate SNR asymptotic is
actually controlled by the continuous part of the pseudo-codeword
spectra above the gap. This prediction is indeed consistent with MC
results shown in the second plot from the top row of
Fig.~(\ref{fig:MC+dens}), see also \cite{06SC}, where the
intermediate asymptotic, $\sim\exp(-27.33 s^2/2)$, set in at
moderate SNRs, changes to a shallower curve with the SNR increase.

The two-stage scenario, when the lowest distance configuration is
the one of a codeword separated by a gap from the rest of the
spectrum, is also seen in the frequency spectra of the
$[648,324,15]$ and $[648,432,12]$ codes, shown in the third row of
the Fig.~(\ref{fig:MC+dens}). However the gaps in the later cases
are much smaller than in the $[672,336,16]$ case. The behavior of
the $[273,191,18]$ code can also be attributed to the same type,
with the exception of one really special feature of the code. Here
one gets a whole stairway of low distance codewords observed with
significant frequencies. Note that the original projective geometry
code has highly connected checks, with degree $17$,  thus the
aforementioned analysis is feasible only for the dendro version of
the code.

Analyzing the $[1296,648]$ code, one finds that, on the one hand,
the configuration with the lowest effective distance,
$d_{min}\approx 19.6$, is a non-codeword pseudo-codeword, like in
the case of the $[155,64,20]$ code. On the other hand, this lowest
configuration is separated by a noticeable gap from the next one
with $d\approx 21.75$, like in the case of the $[672,336,16]$ code.
However, the lowest weight configuration is not rare,  thus
suggesting that the FER vs SNR dependence for this code will likely
show a one stage error-floor associated with the lowest effective
distance.

This work was carried out under the auspices of the National Nuclear
Security Administration of the U.S. Department of Energy at Los
Alamos National Laboratory under Contract No. DE-AC52-06NA25396.

\end{document}